\documentclass[journal]{IEEEtran}

\usepackage{cite}
\usepackage{amsmath}
\usepackage{amsthm}
\usepackage[pdftex]{graphicx}

\newtheorem{thm}{Theorem}
\newtheorem{lemma}{Lemma}

\begin{document}

\title{A Note on Slepian-Wolf Bounds for Several Node Grouping Configurations}

\author{Benjamin~Rosen, Adnan~M.~Abu-Mahfouz,~\IEEEmembership{Senior Member,~IEEE,} and Ling~Cheng,~\IEEEmembership{Senior Member,~IEEE}%
\thanks{Benjamin Rosen and Ling Cheng are with the School of Electrical and Information Engineering, University of the Witwatersrand, Johannesburg, South Africa, e-mail: ling.cheng@wits.ac.za.}% 
\thanks{A. M. Mahfouz is with the Council for Scientific and Industrial Research (CSIR), Pretoria, South Africa}}

\maketitle

\begin{abstract}
The Slepian-Wolf bound on the admissible coding rate forms the most fundamental aspect of distributed source coding. As such, it is necessary to provide a framework with which to model more practical scenarios with respect to the arrangement of nodes in order to make Slepian-Wolf coding more suitable for multi-node Wireless Sensor Networks. This paper provides two practical scenarios in order to achieve this aim. The first is by grouping the nodes based on correlation while the second involves simplifying the structure using Markov correlation. It is found that although the bounds of these scenarios are more restrictive than the original Slepian-Wolf bound, the overall model and bound are simplified.
\end{abstract}


\section{Introduction}

\IEEEPARstart{W}{ith} the massive improvements in processing power and reduction in cost, modern computers are now more suited than ever to be used in applications requiring a lot of data, such as Wireless Sensor Networks (WSNs), in which many nodes observe an event and communicate with a central node. This setup necessitates the implementation of advanced coding techniques to maximise the data rate while minimising the interference amongst nodes. Slepian-Wolf (SW) coding~\cite{1055037} has proven itself as a viable means to performing distributed coding that is suitable for a system of correlated nodes, since WSNs employ many nodes that are in close proximity, meaning that the nodes are highly correlated with one another~\cite{1328091}. Fundamentally, the SW coding rate is bounded, which limits the maximum compression that can be achieved. Although much research has been done on the techniques that achieve this bound~\cite{755665,1281474,1281475,4298397,5339955,7086847,8309135}, the literature is lacking in the adjustment of this bound depending on the nature of the correlation between nodes.

This paper attempts to bridge this knowledge gap by providing the adjusted SW bound for a variety of correlation structures. Specifically, we present a simpler bound than the general SW bound by introducing the concepts of ``strong'' and ``weak'' correlation, and using an arbitrary adjacency matrix to represent the correlation structure, as well as considering nodes in Markov chain relations. We develop this concept to find the bound in a specific case, where the nodes are organised into a disjoint grouping structure. These bounds simplify the modelling for the multi-node bound of the coding rate, while sacrificing the maximum achievability. However, this also has the added effect of reducing the complexity of the coding and decoding schemes. These results are also important with regards to security, since the shared information (given by the bound) results in more information leakage, giving more power to an eavesdropper tapping into the channel \cite{7028606,7282997}.

This paper is organised as follows: Section~\ref{sec:background} gives a brief overview of SW coding, while Section~\ref{sec:sys-model} presents the system model. The total rate bound for different node grouping cases when considering different correlation groupings and Markov chain relations between nodes are discussed in Section~\ref{sec:Ideal} and \ref{sec:constrained}, before concluding in Section~\ref{sec:conclusion}.

\section{Background}\label{sec:background}

In their seminal paper, Slepian and Wolf~\cite{1055037} examined the efficacy of source coding, where the sources are correlated and successive drawings from these sources are i.i.d. Their most notable contribution is the proof that, for a system of two encoders and one decoder, the admissible rate regions are comparable, regardless of whether information sharing between encoders is allowed or not. When the encoders ($x$ and $y$) are allowed to share information, the admissible rate region is bounded by $R_{x}+R_{y}\geq H(X,Y)$. However, if the sharing of information is not allowed, Slepian and Wolf proved that the admissible rate region is given by (\ref{eq:Rx})-(\ref{eq:Rx+Ry}).

\begin{align}
R_{x} & \geq H(X|Y)\label{eq:Rx}\\
R_{y} & \geq H(Y|X)\label{eq:Ry}\\
R_{x}+R_{y} & \geq H(X,Y)\label{eq:Rx+Ry}
\end{align}

\begin{figure*}
    \centering
    \includegraphics{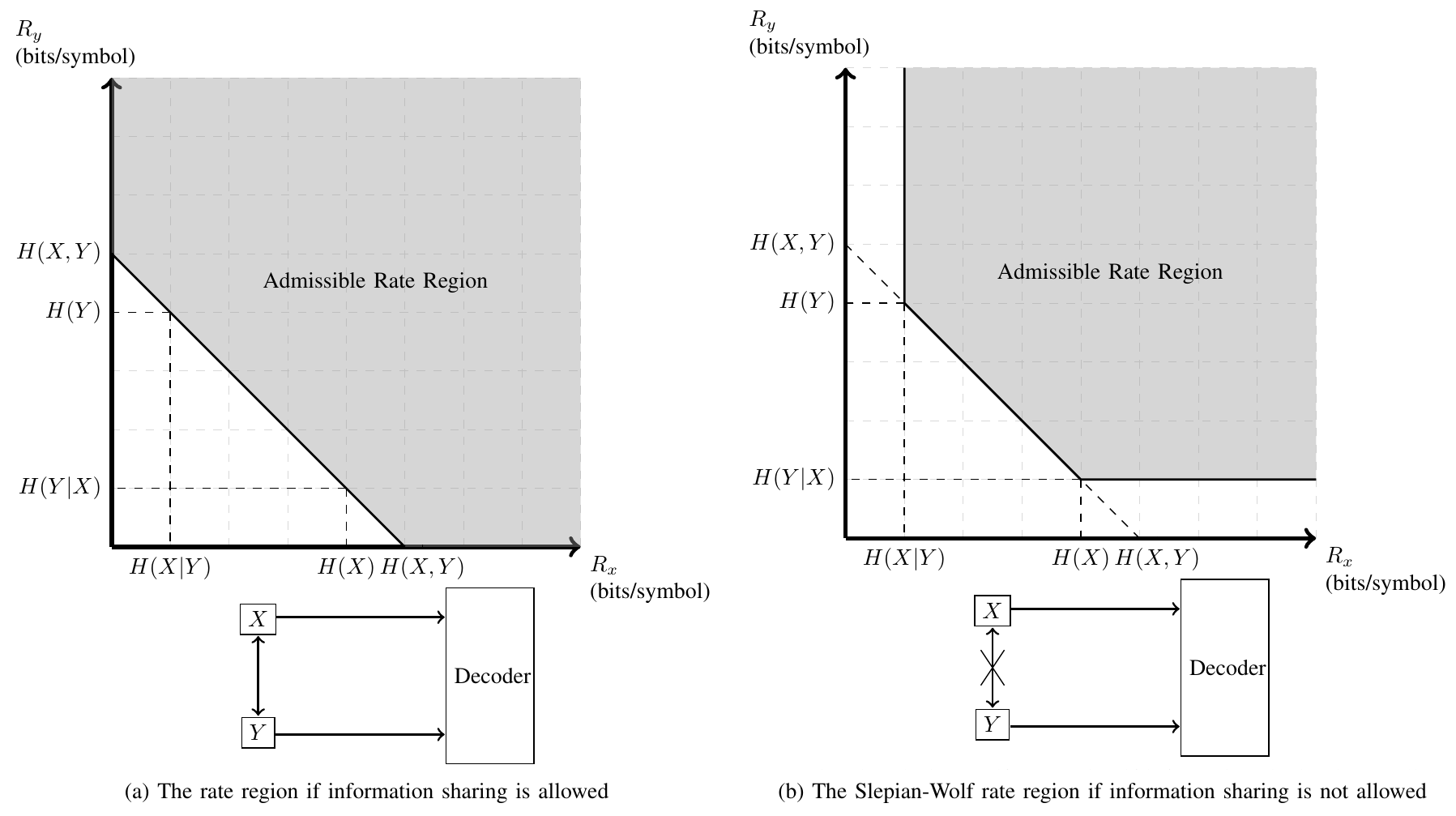}
    \caption{Admissible rate regions for two nodes transmitting to a common receiver}
    \label{fig:Slepian-Wolf-admissible-rate}
\end{figure*}

Fig.~\ref{fig:Slepian-Wolf-admissible-rate} shows that, although the admissible rate region is greater in the former case, it only differs by the maximum achievable rates. This paper has inspired an entire field dedicated to lossless Distributed Source Coding (DSC).\@

Cover~\cite{1055356} extended this contribution by first proving the Slepian-Wolf rate region in a simpler manner. He then extends their work to an arbitrary $N$ number of sources. Of importance to this research is the fact that $N$ sources corresponds to $2^{N}-1$ rate region equations, meaning the maximum admissible rate region can be accurately obtained for any number of sources. Cover gives the following equations for $N=3$, for example:

\begin{equation}
\begin{aligned}R_{1} & \geq H(X_{1}|X_{2},X_{3})\\
R_{2} & \geq H(X_{2}|X_{1},X_{3})\\
R_{3} & \geq H(X_{3}|X_{1},X_{2})\\
R_{1}+R_{2} & \geq H(X_{1},X_{2}|X_{3})\\
R_{2}+R_{3} & \geq H(X_{2},X_{3}|X_{1})\\
R_{1}+R_{3} & \geq H(X_{1},X_{3}|X_{2})\\
R_{1}+R_{2}+R_{3} & \geq H(X_{1},X_{2},X_{3})
\end{aligned}
\label{eq:multi-node}
\end{equation}

\section{System Model}\label{sec:sys-model}
\begin{figure}[ht]
    \centering
    \includegraphics[width=0.35\textwidth]{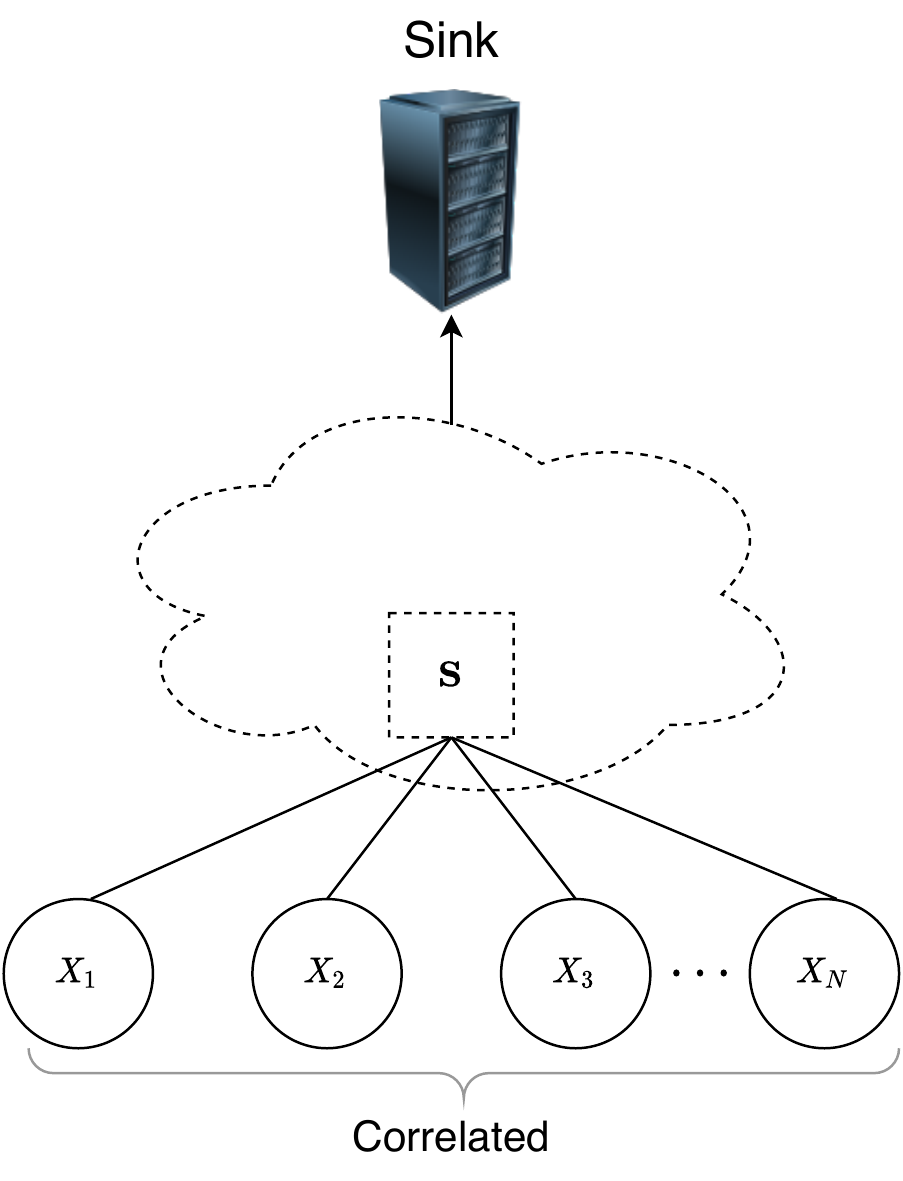}
    \caption{System model of multiple correlated nodes communicating with one receiver}
    \label{fig:system-model}
\end{figure}

The system to be modelled is a part of a WSN, where the edge nodes are correlated with one another. Although they communicate through the WSN, there is only one receiver at the end of the network, denoted by the Sink. Fig.~\ref{fig:system-model} shows the modelling for this system, where $X_{1},\ldots,X_{N}$ are the correlated edge nodes. This paper is particularly focused on the minimum encoding rate over all the nodes, represented by the supernode $\mathbf{S}$.

The grouping of nodes, and thus the bound on the total compression rate, are directly affected by the correlation between nodes. As mentioned in~\cite{1328091}, the correlation represents a ``virtual'' channel that allows the nodes to communicate, although explicit communication is not present. Therefore, this paper focuses on the different arrangement of nodes, based on different correlation scenarios, and how it affects the total rate. We also consider the effect on the total rate when taking into account a Markov chain correlation between nodes.

\section{Total Rate for Single Grouping Cases\label{sec:Ideal}}
In the Section~\ref{sec:background}, the total rate was given for the two and three node case, where all nodes are correlated with one another. We now give a simplified form of Cover's~\cite{1055356} total rate when considering $N$ nodes.

\begin{thm}
For a system of $N$ correlated sources $X_{1}, X_{2}, \ldots X_{N}$, a valid total rate $R_{T}=\sum_{i=1}^{N}R_{i}$ (in bits/symbol) is given by\label{thm:ideal-bound}
\begin{equation}
R_{T}\geq H(X_{1})+\sum_{i=2}^{N}H(X_{i}|X_{i-1},\ldots,X_{1})\label{eq:total_rate_SW}
\end{equation}
\end{thm}

\begin{IEEEproof}
We begin with the final inequality given by Cover in~\cite{1055356} for $N$ nodes:

\begin{equation}
\sum_{i=1}^{N}R_{i}\geq H(X_{1},X_{2},\ldots,X_{N})
\end{equation}
Using the chain rule for entropy and substituting $R_{T}$ yields

\begin{equation}
R_{T}\geq H(X_{1})+H(X_{2}|X_{1})+\cdots+H(X_{N}|X_{N-1},\ldots,X_{1})
\end{equation}
which can be simplified using a summation.
\end{IEEEproof}

Conceptually, the total rate given in Theorem~\ref{thm:ideal-bound} represents a single valid point on the $N-1$ hypersurface defining the upper bound. Any point on this surface will have the same total rate, with a different distribution of individual rates. The result shown here is a parallel to the ``corner points'' given in literature for the two node case~\cite{4518272}.

\begin{figure}[ht]
    \centering
    \includegraphics[width=0.4\textwidth]{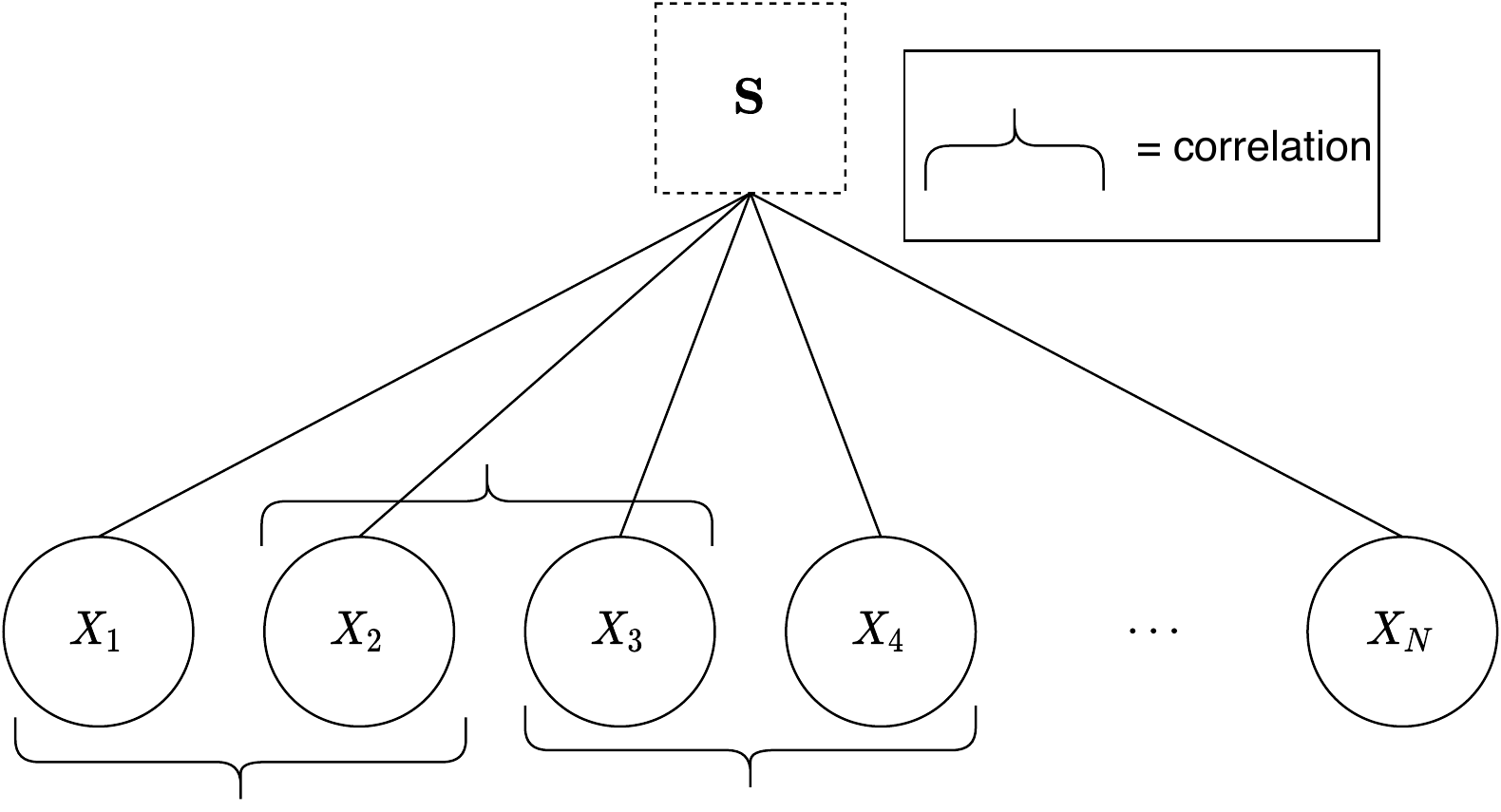}
    \caption{Modelling a DSC system with Markov correlation}
    \label{fig:markov-correlation}
\end{figure}

Next we consider the lemma of this bound calculation if the correlation is modelled as a Markov chain. This implies that each node is correlated with its direct neighbours, but is independent of all other nodes (as shown in Fig.~\ref{fig:markov-correlation}).

\begin{lemma}
For a system of $N$ nodes, where the correlation between them forms a Markov chain $X_{1}\rightarrow X_{2}\rightarrow\cdots\rightarrow X_{N}$, a valid bound on the total rate $R_{T}$ is given by\label{lem:markov-correlation}
\begin{equation}
R_{T}\geq H(X_{1})+\sum_{i=2}^{N}H(X_{i}|X_{i-1})\label{eq:markov-bound}
\end{equation}
\end{lemma}

\begin{IEEEproof}
From Theorem~\ref{thm:ideal-bound} we have
\begin{equation}
R_{T}\geq H(X_{1})+H(X_{2}|X_{1})+\cdots+H(X_{N}|X_{N-1},\ldots,X_{1})\label{eq:all_nodes_bound}
\end{equation}
By the definition of Markov correlation, $H(X_{3}|X_{2},X_{1})=H(X_{3}|X_{2})$ and $H(X_{4}|X_{3},X_{2},X_{1})=H(X_{4}|X_{3})$. Thus, every term in~\eqref{eq:all_nodes_bound} to the right of the inequality (besides for the first) will reduce to $H(X_{i}|X_{i-1})$.
\end{IEEEproof}

We can also consider a system in which there is a combination of Markov and regular correlation structures.

\begin{thm}
For a system of $N$ sources, correlated by a combined regular and Markov correlation $X_{1},\ldots,X_{k-1}\rightarrow X_{k}\rightarrow\cdots\rightarrow X_{k+r},X_{k+r+1},\ldots,X_{N}$, $k+r+1<N$, the bound on the total rate is\label{thm:markov-ideal}

\begin{multline}
R_{T}\ge H(X_{1})+\sum_{i=2}^{k}H(X_{i}|X_{i-1},\ldots,X_{1})\\
+\sum_{j=k+1}^{k+r+1}H(X_{j}|X_{j-1})+\sum_{i=k+r+2}^{N}H(X_{i}|X_{i-1},\ldots,X_{k+r})
\end{multline}
\end{thm}

\begin{IEEEproof}
The bound follows Theorem~\ref{thm:ideal-bound} for $X_{i}$, $i=1,\ldots,k$ and Theorem~\ref{thm:markov-ideal} for $i=k+1,\ldots,k+r+1$, while for $k+r+1<i<N$, Theorem~\ref{thm:ideal-bound} holds again, with $X_{k+r}$ replacing $X_{1}$ since $H(X_{i}|X_{i-1},\ldots,X_{1})=H(X_{i}|X_{i-1},\ldots,X_{k+r})$ and the Markov independence breaks the entropy chain.
\end{IEEEproof}

\section{Total Rate for Categorised Correlation Cases}\label{sec:constrained}

In order to simplify the SW bound, it is necessary to organise the nodes into different arrangements based on more practical correlation aspects. Although there are many methods to quantifying the correlation between two entities, we use a correlation function $f_{C}(X_{1},X_{2})$, which evaluates the correlation between two nodes $X_{1}$ and $X_{2}$ by using a value (distance) metric, in which $\tau$ is the threshold, such that $f_{C}(X_{1},X_{2})\geq\tau$ means the correlation between the nodes is strong, with weak correlation otherwise. We use a graph to represent the inter-correlation between $N$ nodes, which can be represented using an adjacency matrix $\boldsymbol{C}$, where $c_{i,j}=f_{C}(X_{i},X_{j})$ and $c_{i,j}=0$ if the correlation is weak, or if $i=j$. We can construct a set for each row $i$ of $\boldsymbol{C}$, defined as $S_{i}=\{j|j=1, 2, \ldots,i-1;c_{i,j}\neq0\}$, which contains all the indices of nodes correlated with $X_{i}$, where $j<i$. Furthermore, we use the notation $\mathcal{X}_{S_{i}}$ to represent all nodes referenced by the set $S_{i}$, namely $X_{j_{1}},X_{j_{2}},\ldots,X_{j_{|S_{i}|}}$. We now give the SW bound when using this correlation structure.

\begin{thm}
Given an adjacency matrix $\boldsymbol{C}$ containing the correlation information for $X_{1},X_{2},\ldots,X_{N}$ the bound on the admissible total rate is given by

\begin{equation}
R_{T}\geq H(X_{1})+\sum_{i=2}^{N}H(X_{i}|\mathcal{X}_{S_{i}})
\end{equation}
\end{thm}

\begin{IEEEproof}
The definition of $S_{i}$ is designed to capture the lower triangular part of $\boldsymbol{C}$. This definition allows the same structure as that in Theorem~\ref{thm:ideal-bound} to be followed, which is the expansion of $H(X_1,X_2,\ldots,X_N)$. The difference lies in the fact that $\mathcal{X}_{S_{i}}$ excludes any terms that are not correlated with $X_{i}$, giving the desired expansion.
\end{IEEEproof}

\begin{figure}[ht]
    \centering
    \includegraphics[width=0.45\textwidth]{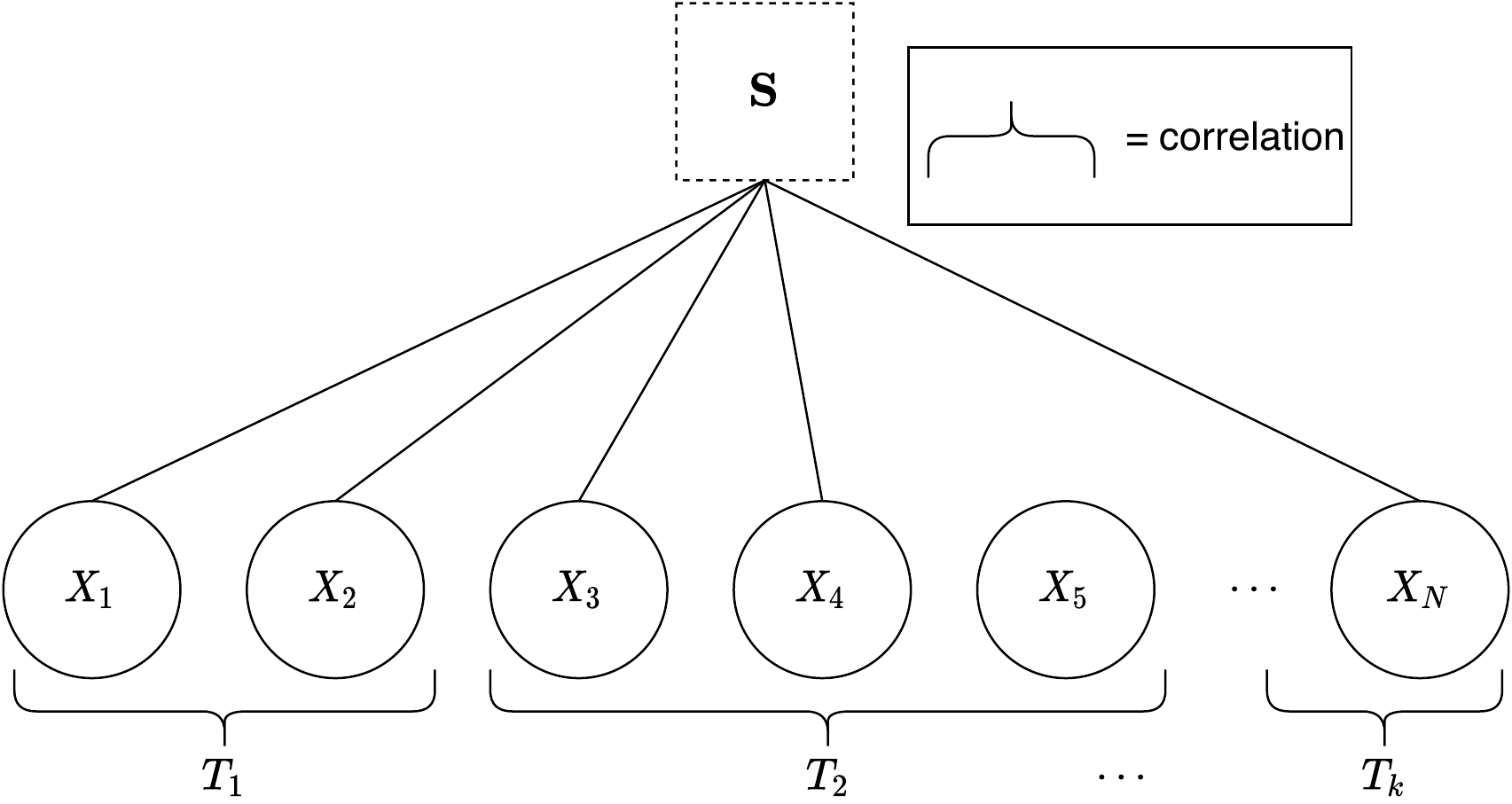}
    \caption{Modelling a system with disjoint correlation}
    \label{fig:disjoint-correlation}
\end{figure}

In the next scenario, we fix $\tau$ such that $\mathbf{C}$ becomes sparse, implying that only specific subsets of nodes are strongly correlated with one another, but not with any other group of nodes. The modelling for this disjoint grouping case is shown in Fig.~\ref{fig:disjoint-correlation}. Clearly, the set of all nodes is divided into $k$ independent sets. Let $T_{i}$ be a set containing the indices of the nodes in the $i\mathrm{th}$ independent set, with $i=1,2,\ldots,k$. Necessarily, $T_{p}\cap T_{q}=\emptyset$ $\forall p,q\in\{1,2,\ldots,k\},p\neq q$. Furthermore, let $X_{i,j}$ be the node referenced by the $j\mathrm{th}$ element of $T_{i}$, with $j=1, 2, \ldots,|T_{i}|$.

\begin{thm}
For a system of $N$ correlated sources $X_{1}, X_{2}, \ldots,X_{N}$, and disjoint set distribution $T_{i}$, $i=1, 2, \ldots,k$, a valid total rate is given by\label{thm:disjoint-bound}

\begin{equation}
R_{T}\geq\sum_{i=1}^{k}H(X_{i,1})\sum_{j=2}^{|T_{i}|}H(X_{i,j}|X_{i,j-1},\ldots,X_{i,1})
\end{equation}
\end{thm}

\begin{IEEEproof}
For each $T_{i}$, the total rate is given by Theorem~\ref{thm:ideal-bound}, replacing $X_{1},X_{2},\ldots,X_{N}$ with $X_{i,1},X_{i,2},\ldots,X_{i,|T_{i}|}$ accordingly:

\begin{multline}
   R_{i,1}+R_{i,2}+\cdots+R_{i,|T_{i}|}\geq H(X_{i,1})\\
+\sum_{j=2}^{|T_{i}|}H(X_{i,j}|X_{i,j-1},\ldots,X_{i,1}) 
\end{multline}

Since each subset is independent, the bound for each total rate associated with $T_{i}$ can be added without any further manipulations.
\end{IEEEproof}

As in the Section~\ref{sec:Ideal}, we can also determine the adjusted bound when the nodes within each disjoint set are correlated with a Markov relation.

\begin{lemma}
For a system of $k$ disjointly correlated nodes represented by $T_{i}$ and modelled with a Markov correlation between them ($X_{i,1}\rightarrow X_{i,2}\rightarrow\cdots\rightarrow X_{i,|T_{i}|}$, $i=1,2,\ldots,k$), a valid total rate is given by\label{thm:disjoint-markov-bound}

\begin{equation}
    R_{T}\geq\sum_{i=1}^{k}H(X_{i,1})\sum_{j=2}^{|T_{i}|}H(X_{i,j}|X_{i,j-1})
\end{equation}
\end{lemma}

\begin{IEEEproof}
    The proof follows along the same line of reasoning as Theorem~\ref{thm:disjoint-bound}, using Lemma~\ref{lem:markov-correlation} instead of Theorem~\ref{thm:ideal-bound}.
\end{IEEEproof}

\section{Conclusion}\label{sec:conclusion}

The Slepian-Wolf bound has been shown in both its original form, as well as the simplification when modelling the correlation between nodes as a Markov chain. These bounds have then been extended to include an arbitrary correlation structure, by introducing and defining the concepts of ``strong'' and ``weak'' correlation and using an adjacency matrix to model the correlation. We also consider the effect on the bound when modelling the correlation with a Markov chain structure. A final simplification is demonstrated by making the adjacency matrix sparse, causing the nodes to be arranged into disjoint groups. Although the bounds presented in this paper are more restrictive than the Slepian-Wolf bound, the modelling complexity is reduced owing to the more practical correlation considerations. By outlining a simple yet efficient modelling approach, we show the potential of improving the coding/decoding complexity of practical SW algorithms and schemes.

\section*{Acknowledgement}

This research was partially supported by the Council for Scientific and Industrial Research, Pretoria, South Africa, through the Smart Networks collaboration initiative and IoT-Factory Program (Funded by the Department of Science and Innovation (DSI), South Africa), and partially by South Africa’s National Research Foundation (129311)

\bibliographystyle{IEEEtran}
\bibliography{References}


%\begin{IEEEbiography}{Benjamin Rosen}
%\end{IEEEbiography}

%\begin{IEEEbiographynophoto}{Adnan M. Abu-Mahfouz}

%\newpage

%\begin{IEEEbiographynophoto}{Ling Cheng}


%\vfill


\end{document}